\documentclass[10pt,conference]{IEEEtran}

\usepackage[utf8]{inputenc}

\usepackage[textwidth=16cm,textheight=21.7cm]{geometry}

\usepackage{graphicx}
\usepackage{physics}
\usepackage{amsmath}
\usepackage{amssymb}
\usepackage{bbm}
\usepackage{subcaption}
\usepackage{cite}
\usepackage{authblk}

\usepackage{tikz}
\usetikzlibrary{automata, arrows, positioning}

\usepackage{tikz-network}

\usepackage{acronym}
\newacro{CNOT}{controlled $X$}
\newacro{GHZ}{Greenberger-Horne-Zeillinger}

\title{A quantum walk control plane for distributed quantum computing in quantum networks}
\author[1]{Matheus Guedes de Andrade}
\author[1, 2]{Wenhan Dai}
\author[3]{Saikat Guha}
\author[1]{Don Towsley}
\affil[1]{College of Information and Computer Science, University of Massachusetts Amherst}
\affil[2]{Quantum Photonics Laboratory, Massachusetts Institute of Technology}
\affil[3]{College of Optical Sciences, University of Arizona}

\date{November 25, 2020}

\begin{document}

\maketitle

\begin{abstract}
    Quantum networks are complex systems formed by the interaction among quantum processors through quantum channels. Analogous to classical computer networks, quantum networks allow for the distribution of quantum computation among quantum computers. In this work, we describe a quantum walk protocol to perform distributed quantum computing in a quantum network. The protocol uses a quantum walk as a quantum control signal to perform distributed quantum operations. We consider a generalization of the discrete-time coined quantum walk model that accounts for the interaction between a quantum walker system in the network graph with quantum registers inside the network nodes. The protocol logically captures distributed quantum computing, abstracting hardware implementation and the transmission of quantum information through channels. Control signal transmission is mapped to the propagation of the walker system across the network, while interactions between the control layer and the quantum registers are embedded into the application of coin operators. We demonstrate how to use the quantum walker system to perform a distributed CNOT operation, which shows the universality of the protocol for distributed quantum computing. Furthermore, we apply the protocol to the task of entanglement distribution in a quantum network.
\end{abstract}

\section{Introduction}

Quantum networking is an innovative, multidisciplinary field of research that promises revolutionary improvements in communications, enabling tasks and applications that are impossible to achieve with the exclusive exchange of classical information \cite{kimble2008quantum, wehner2018quantum}. Similar to a classical computer network, a quantum network is a distributed system composed of quantum computers and quantum repeaters that exchange quantum information across physical channels. Among applications supported by quantum networks, distributed quantum computing is of particular interest as it leverages the power of interconnected quantum computers to create a virtual quantum machine with processing capabilities that surpass its physical constituents alone \cite{beals2013efficient, cacciapuoti2019quantum, gyongyosi2021scalable}. Distributed quantum computing becomes even more interesting in the noise intermediate scale quantum machines (NISQ) scenario where there is a clear tradeoff between the size of quantum computers, in terms of number of qubits, and the fidelity of quantum operations, given the fact that physical separation directly reduces cross talk among qubits \cite{van2016path}. When the quantum network scenario is considered, the complexity of distributed quantum computing extends in at least two dimensions. First, physical quantum channels have a well known depleting effect in the exchange of quantum data, \textit{e.g.}, the exponential decrease in channel entanglement rate with distance~\cite{pirandola2017fundamental}. Second, there is a demand for a quantum network protocol capable of performing a desired distributed quantum operation while accounting for network connectivity. Generic quantum computation with qubits in distinct quantum processors demands either the application of remote controlled gates~\cite{chou2018deterministic} or the incessant exchange of quantum information.
For both cases, a network protocol is necessary to orchestrate the communication between nodes that are not directly connected with one another.


One challenge in the design of a control protocol is the need for being agnostic to hardware implementations. There is a plethora of physical systems suited for quantum computation under investigation, \textit{e.g.},  superconducting qubits~\cite{arute2019quantum}, trapped ions~\cite{kielpinski2002architecture, pino2021demonstration} and Silicon-vacancy color centers in diamond~\cite{stas2021high, wang2021robust}. In addition, there is a diverse investigation in the architectural description of quantum interconnecting devices capable of exchanging quantum information encoded in distinct quantum physical quantities~\cite{awschalom2021development}. This diverse ecosystem of quantum network technologies indicates that distributed quantum computing network protocols need to abstract physical implementations of quantum switches and network connectivity while maintaining universality requirements.

The goal of this article is to propose a control protocol for distributed quantum computing based on discrete time quantum walks \cite{aharonov2001quantum}. Quantum walks are universal for quantum computing \cite{childs2009universal, childs2013universal, lovett2010universal} and have been successfully employed in the quantum network scenario to perform perfect state transfer (PST) between network nodes \cite{nitsche2016walk, zhan2014perfect}, teleportation \cite{li2019new, wang2017generalized} and quantum key distribution (QKD)~\cite{vlachou2018quantum}. Previous works describe ways of distributing entanglement between nodes on a quantum network using the coin space of the walker to propel entanglement generation between qubits \cite{li2019new, zhan2014perfect}. In addition, the formalism of quantum walks with multiple coins enabled the description of an entanglement routing protocol, which interprets qubits within a network node as vertices of an abstract graph used by a quantum walker to generate entanglement~\cite{li2020entangled}. In spite of their relevance, the quantum walk approaches defined in the literature are suited to particular network structures and quantum operations. In particular, they consider the case of regular lattices, describing walker dynamics on regular structures and do not address how the quantum walk can be used to perform generic distributed quantum operators. In this context, this work adds to the literature of both quantum walks and quantum networks with the description of a quantum network control protocol that can be applied to arbitrary graphs and perform universal quantum computing in a quantum network.


\subsection{Contributions}

The contributions of this article are three-fold.
\begin{itemize}
    \item We propose a quantum walk protocol for distributed quantum computing in a quantum network. The protocol uses a quantum walker system as a quantum control signal to perform computations among quantum processors that are physically separated. We assume that each processor dedicates part of its internal quantum register to represent the walker control signal and describe how the control subsystem interacts with the data subsystem. The interaction between data and control is specified by unitary operations that nodes need to implement in order to realize the quantum walk control plane.
    \item We show the universality of the protocol by describing how a 2-qubit \ac{CNOT} operation between qubits in distinct nodes of the network can be performed with the quantum walk. The protocol is universal in the sense that it allows for any quantum operation in the Hilbert space formed by all qubits in the data subsystem of the nodes. Furthermore, it is generic in the sense that it abstracts hardware implementation and channel transmissions, while being well-defined for any network topology.
    \item We demonstrate how the protocol can be used to recover the behavior of entanglement distribution protocols previously described in the literature~\cite{pant2019routing, patil2020entanglement}.
\end{itemize}
 
The description of the protocol is carried in the logical setting under the assumption that quantum error correction processes ensure unitary operations for both control and data qubits.

The remainder of this article is structured as follows. In Section 2, we present the mathematical background needed for the description of the quantum walk protocol. We describe the quantum walk protocol and demonstrate its universality in Section 3. The description is extended to case of multiple walkers in Section 4. In Section 5, we apply the protocol to recover the behavior of entanglement distribution protocols. Finally, the manuscript is concluded in Section 6.

\section{Background}

Consider the graph $G = (V, E)$. Let $\delta(v)$ denote the set of neighbors of node $v \in V$ and $d(v) = |\delta(v)|$ denote the degree of $v$. Let $\Delta(u, v)$ denote the hop-distance between $u$ and $v$ in $G$. Throughout this work we refer to the inverse of a binary string $x \in \{0, 1\}^{*}$ as $\overline{x}$.
A quantum network is a set of quantum hosts (quantum processors) interconnected by a set of quantum channels that allow for the exchange of quantum information \cite{wehner2018quantum}. A host is either a quantum repeater, a quantum router or a quantum computer with a fixed number of qubits, which performs generic quantum operations. A quantum network can be represented as a symmetric directed graph $N = (V, E)$. Each node $v \in V$ represents a quantum host that has a set $\mathcal{M}_v$ of qubits that can be processed together at any time and a set $\mathcal{N}_v$ of qubits that is used to exchange quantum information with nodes in its neighborhood $\delta(v)$ through a set of quantum channels. More precisely, each edge $(u, v) \in E$ represents a quantum channel connecting the qubits in $\mathcal{N}_u$ and $\mathcal{N}_v$ which can interact through operations mediated by the channel. We will refer to $\mathcal{N}_v$ and $\mathcal{M}_v$ as the control and data registers of node $v$, respectively. The sets $\mathcal{M} = \bigcup_v \mathcal{M}_v$ and $\mathcal{N} = \bigcup_v \mathcal{N}_v$ are respectively referred to as the \textit{network control plane} and the \textit{network data plane}.

This network model separates the qubit registers in the nodes into control and data registers. This choice differs from previous works where the set of qubits $\mathcal{N}$ that can interact with quantum channels are considered alone~\cite{pant2019routing, patil2020entanglement}. Note that considering only control qubits suffices to define entanglement routing protocols. Nonetheless, it is straightforward for the goal of describing a quantum network control plane protocol to consider the separation between control and data registers in the nodes.


\subsection{Quantum network protocols}

Consider the system formed by two quantum processors $u$ and $v$ connected by a channel and their respective qubits. A \textit{local operation} is a quantum transformation represented as a separable operator of the form
\begin{align}
    & O = O_u \otimes O_v,
\end{align}
where $O_u$ and $O_v$ act on the state space of the qubits at processors $u$ and $v$, respectively. A \textit{local operation assisted by two-way classical communication} (LOCC) is a local operation that depends on classical information exchanged between nodes \cite{pirandola2017fundamental}, \textit{e.g.,} the unitaries of quantum teleportation. The classical information is used to select operators applied to the system, which is embedded in the formalism by averaging non-unitary measurement operators into trace-preserving operators. 
A quantum network protocol for $N$ is an algorithm that operates on the qubits at the nodes, transforming their joint state by an LOCC $\bar{\Lambda}_t$ as
\begin{align}
    \rho(t) = \bar{\Lambda}_t \rho(0) \bar{\Lambda}_t^{\dagger},
\end{align}
where $t$ is the number of rounds in the protocol, $\rho$ is a density operator defined on the Hilbert space $\mathcal{H}_{\mathcal{M}\cup \mathcal{N}}$ formed by all qubits on the network and $\rho(0)$ is a density matrix where registers in distinct nodes are in a separable state. We express $\bar{\Lambda}_t = \prod_{k = 0}^{t}\Lambda_k$ as a product of other LOCCs each applied at different time steps~\cite{pirandola2019end}. Note that it suffices for $\rho(0)$ to be a separable density operator in~\cite{pirandola2019end}. In fact, it is also possible to model a protocol as a sequence of external (mediated by channels) and internal (local to vertices) time-dependent superoperators $A_t$ and $B_t$ such that the state of the network is described as
\begin{align}
    & \rho(t + 1) = B_{t}[A_{t}[\rho(t)]],\label{eq:protocol_model}
\end{align}
where $\rho$ is a density matrix characterizing the joint state of all network memories at discrete time $t$. As an example, it is usually the case for entanglement distribution protocols that $A_t$ represents channel entanglement protocols performed in all channels of the network and $B_t$ represents either entanglement swapping operations or GHZ projections performed independently in multiple nodes. The LOCC formalism has proven useful since it allowed for the derivation of fundamental bounds for entanglement rate distribution \cite{pirandola2019end}. In this article, however, our focus is to address network protocols for the transmission of quantum information in the logical perspective. In particular, we exploit the representation in \eqref{eq:protocol_model} considering $A$ and $B$ as unitary operators provided by the network instead of generic superoperators and perform the analysis in the state vector formalism. It is worth emphasizing that the unitarity assumption for operators $A$ and $B$ translates to the assumption that quantum error correction is provided by the network.

\subsection{Quantum walks on graphs}

There are many ways to define a quantum walk on a graph and this article focuses on the discrete-time coined quantum walk model. Given a symmetric directed graph $G = (V, E)$, a coined quantum walk on $G$ is a process of unitary evolution on the Hilbert space $\mathcal{H}_G = \mathcal{H}_{V} \otimes \mathcal{H}_\mathcal{C}$ formed by the edges of the graph, where $\mathcal{H}_{V}$ codifies vertices and $\mathcal{H}_\mathcal{C}$ is the coin space of the walker codifying the degrees of freedom the walker can move on. More precisely, every $(v, u) \in E$ defines a basis vector $\ket{v, c}$ for $\mathcal{H}_G$ through a mapping between the edges incident to $v$ with the set of degrees of freedom $\mathcal{C}_v = \{0, 1, \ldots, d(v) - 1\}$. As a convention, we map the degrees of freedom of the walker at a given node $v$ to the order of labels of its neighbors, such that $\ket{v, c}$ refers to the edge $(v, u)$, where $u$ is the $c$-th smallest label in $\delta(v)$. Later, $c_{vu}$ will be used to refer to the degree of freedom of $v$ that represents the edge $(v, u)$ and $c_{v}$ to represent the self-loop $(v, v)$. The generic state of the walker $\ket{\Psi(t)} = \sum \psi(v, c, t) \ket{v, c}$ is a superposition of the edges of $G$ and the walker evolution is defined as
\begin{align}
    & \ket{\Psi(t + 1)} = S(t)C(t) \ket{\Psi(t)} \label{eq:qwalk},
\end{align}
where $C$ and $S$ are respectively referred to as the coin and shift operators. The coin is a unitary operator of the form
\begin{align}
    & C(t) = \sum_{v} \dyad{v} \otimes C_v(t), \label{eq:walk_coin}
\end{align}
where $C_v: \mathcal{H}_{\mathcal{C}} \to \mathcal{H}_{\mathcal{C}}$.
The shift can be defined as any permutation operator on the edges of the graph that maps an edge $(v_1, u)$ to an edge $(u, v_2)$. This mapping of edges represents a permutation between states $\ket{v_1, c}$ and $\ket{u, c'}$, where $u$ is the $c$-th neighbor of $v_1$. Two shift operators are used throughout this work: the identity operator, which is a trivial permutation of the edges, and the flip-flop shift operator given by
\begin{align}
    & S_f = \sum_{v \in V}\sum_{u \in \delta(v)} \dyad*{v, c_{vu}}{u, c_{uv}} \label{eq:flipflop}, 
\end{align}
which applies, for every $(v, u) \in E$, the permutation $(v, u) \to (u, v)$. $S_f$ reverses edges in the walker wavefunction and is well defined for every symmetric directed graph. The label of nodes and degrees of freedom are numerical and expressed as binary strings. In this setting, the degree of freedom $\overline{c}$ represents the bit-wise negation of label $c$. Note that $c_{vu}$ is not necessarily $\overline{c_{uv}}$.

\section{Quantum walk network control plane}\label{sec:assisted}

A direct way of controlling operations in a quantum network with a quantum walk is to consider a coupled system between a walker and the qubits in the nodes, using supersposition in the walker system to implement controlled unitary operations. We describe this joint system in the logical setting. Let $N' = (V, E  \cup \{(v, v), \forall v \in V\})$ be the graph obtained by adding self-loops to a network $N$. $\mathcal{H}_W = \mathcal{H}_V \otimes \mathcal{H}_{\mathcal{C}}$ denotes the space of a walker system on $N'$ 
and $\mathcal{H}_g = \mathcal{H}_W \otimes \mathcal{H}_\mathcal{M}$ denotes the global Hilbert space spanned by the coupled systems. The walker system is assumed to be implemented in the networking control plane, such that $\mathcal{H}_W$ is a subspace of $\mathcal{H}_\mathcal{N}$. Since the Hilbert space of the walker system represents an edge of $N'$ as a basis vector, the dimension of the Hilbert space $\mathcal{H}_\mathcal{N}$ must be at least $|E|$ for $\mathcal{H}_\mathcal{W}$ to be a subspace. We consider $\mathcal{H}_g$ as a subspace of $\mathcal{H}_\mathcal{N} \otimes \mathcal{H}_\mathcal{M}$.

\subsection{Control-data interactions}

We prescribe two laws of interaction between the quantum walk system and the data qubits in the network. The first generalizes a coin operator to the global Hilbert space $\mathcal{H}_g$ following
\begin{align}
    & C(t) = \sum_{v} \dyad*{v} \otimes C_v(t) \otimes U_{v}(t) \label{eq:coin_op_assited},
\end{align}
where $C_v(t): \mathcal{H}_{\mathcal{C}} \to \mathcal{H}_{\mathcal{C}} $ is a unitary operator on the coin space of the walker and $U_v(t): \mathcal{H}_\mathcal{M} \to \mathcal{H}_\mathcal{M}$ is a unitary operator on the data space of the network written as
\begin{align}
    & U_v(t) = K_v(t) \bigotimes_{u \neq v} I_{\mathcal{M}_v}, \label{eq:unit_expanded}
\end{align}
where $I_{\mathcal{M}_v}$ is the identity operator for the Hilbert space spanned by all qubits of node $v$.
Essentially, $C(t)$ applies the operator $K_u(t)$ on qubits in $v$ if, and only if, the walker has a non-zero wavefunction component in $v$. Note that $U_v(t)$ is the extension of $K_v(t)$, which acts in $\mathcal{H}_{\mathcal{M}_v}$, to $\mathcal{H}_\mathcal{M}$ with the identity operator. Note that the operator described in \eqref{eq:coin_op_assited} extends the definition in \eqref{eq:walk_coin} to perform unitary operations in the data qubit space controlled by node position.

The second law of interaction defines controlled operations between the data qubits and the coin space of the walker. In particular, we consider the case where only a single data qubit in a given node is used to control a unitary operation in the coin space of the walker. Let
\begin{align}
    & U_{vq}(t) = \sum_{s \in \{0, 1\}} U_{vqs}(t) \otimes \dyad{s} \bigotimes_{q' \neq q} I_{q'} \label{eq:qubit_control}
\end{align}
be a unitary operator acting on the coin space of node $v$ controlled by the qubit $q \in \mathcal{M}_v$ defined for the joint space $\mathcal{H}_\mathcal{C} \otimes \mathcal{H}_\mathcal{M}$, where $s$ denotes the state of $q$. The complete interaction has the form
\begin{align}
    & O(t) = \sum_{v} \dyad{v} \otimes U_{vq}(t). \label{eq:data_control}
\end{align}

The operators correspond to a distributed implementation of a quantum walk system in the network. In essence, each node $v$ contributes with $\mathcal{N}_v$ to describe the space $\mathcal{H}_W$ in $\mathcal{H}_\mathcal{N}$. Thus, \eqref{eq:coin_op_assited} and \eqref{eq:data_control} are operations controlled by the states of $\mathcal{H}_\mathcal{N}$ that represent walker position since $\ket{v}$ is implemented by an entangled state among qubits in $\mathcal{N}$. This encapsulates the necessary entanglement required for distributed controlled operations between the qubits of the network. As previously mentioned, we assume quantum error correction and describe the walker protocol in terms of logical qubits and operations. In the context of error correcting codes, a single logical qubit is implemented by a set of physical qubits \cite{laflamme1997error, roffe2019quantum}. An implementation of the walker demands $\mathcal{O}(\log(|E|))$ logical qubits in $\mathcal{N}$ since each edge of $N'$ is a basis vector of $\mathcal{H}_W$ and the dimension of the space spanned by qubits is exponential in the number of qubits. For practical implementations, network models that consider physical networking qubits in the order of $\mathcal{O}(|E|)$ have exponentially many physical qubits as the required number of logical qubits for the walker, which is interesting for the purpose of quantum error correction.

\subsection{Universal distributed quantum computing}

The interaction behavior described by \eqref{eq:coin_op_assited} and \eqref{eq:data_control} suffices for universal quantum computing. Our protocol trivially allows for the application of any separable qubit operator in the network nodes, since there is no need for the interaction between data qubits and the walker in this case. Thus, we demonstrate universality by demonstrating how a \ac{CNOT} gate between any pair of qubits $a$ and $b$, respectively at arbitrary nodes $A$ and $B$, can be performed using the quantum walk control plane. In fact, the demonstration works for any two-qubit controlled operation by substituting $X$ with the desired single-qubit operator. Without loss of generality, assume that $a$ is the control qubit. To simplify notation, we only define the operators $C(t)$ in \eqref{eq:coin_op_assited} and $O(t)$ in \eqref{eq:data_control} for the subspaces spanned by the qubits in nodes $A$ and $B$, considering undefined operators to be identities, and omit the qubits in $\mathcal{M} \setminus \{a, b\}$. The subscripts following degrees of freedom inside a ket specifies edges while the ones following qubits are considered as indices, \textit{e.g.,} $\ket*{A, c_A, 0_a, 1_b} \in \mathcal{H}_{W} \otimes \mathcal{H}_4$ represents the walker in the self-loop $(A, A)$, $a$ in the $\ket{0}$ state and $b$ in the $\ket*{1}$ state, where $\mathcal{H}_4$ is the Hilbert space spanned by 2 qubits.

We consider the initial state of the walker to be $\ket*{A, c_{A}}$, which corresponds to the self-loop in $A$. Thus, the global system is described by the state vector
\begin{align}
    & \ket{\Psi(0)} = \ket{A, c_{A}} \otimes (\alpha \ket{0_a} + \beta \ket{1_a}) \otimes \ket{\Psi_b}, \label{eq:separable}
\end{align}
where $\ket{\Psi_b}$ is the state of $b$. The interaction operator $O$ described in \eqref{eq:data_control} is applied with the \ac{CNOT} operation
\begin{align}
    & U_{Aa} = I_{\mathcal{C}_A} \otimes \dyad{0} + X_{\mathcal{C}_A} \otimes \dyad{1},
\end{align}
where $\mathcal{C}_A$ represents the coin space of node $A$ and which trivially generates the entangled state
\begin{align}
    & O \ket{\Psi(0)} = (\alpha \ket{A, c_{A}, 0} + \beta \ket{A, \overline{c_{A}}, 1}) \otimes \ket{\Psi_b}. \label{eq:entanglement_operator}
\end{align}
This entangled state can be propagated in the network through the evolution of the walker system such that the state after propagation becomes
\begin{align}
    & \ket{\Psi(t)} = (\alpha \ket{A,c_{A}, 0} + \beta \ket{B, c_{B}, 1}) \otimes \ket{\Psi_b}. \label{eq:routed}
\end{align}
We refer to the propagation of the walker as the transmission of the control signal through the network, which will be further explained in detail. The last step refers to the application of the extended coin operator defined in \eqref{eq:coin_op_assited}, with $K_b = X$ and all other operators defined as identity. The final state obtained is
\begin{align}
    & \ket*{\Psi(t)} = \alpha \ket{A, c_A, 0, \Psi_b} + \beta \ket{B, c_B, 1, \overline{\Psi}_b} \label{eq:cnot_state},
\end{align}
which clearly shows the application of a CNOT controlled by $a$ with $b$ as a target.

Note that the state obtained in \eqref{eq:cnot_state} is entangled with the walker subsystem. We will later demonstrate how to separate the data qubits from the walker.

\subsection{Protocol execution in the network}

In the context of quantum network protocols, the operators defined by \eqref{eq:coin_op_assited} and \eqref{eq:data_control} do not capture the propagation of control information across the network. The propagation of quantum control information between neighbors in the network is embedded in the formalism through the walker's shift operator. The propagation in the case of non-neighboring nodes needs both coin and shift operators as defined in \eqref{eq:walk_coin} and \eqref{eq:flipflop}. Note that generic computation on the data qubits can be performed with \eqref{eq:coin_op_assited} and \eqref{eq:data_control} given states in the superposition \eqref{eq:cnot_state}, although it is not possible to evolve the walker to that superposition from $\ket*{A, c_A}$ without shift operators.

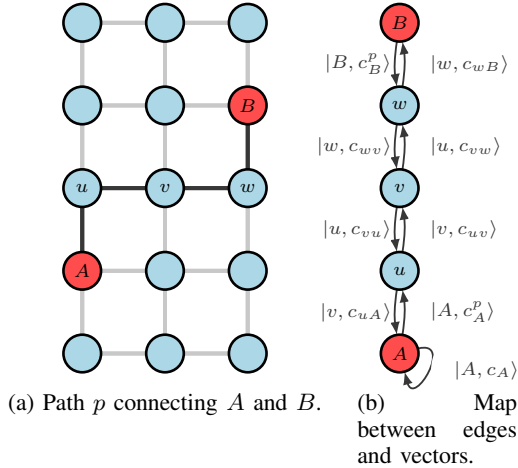
\begin{figure}
    \begin{centering}
        \subfloat[Path $p$ connecting $A$ and $B$.\label{subfig:path_selection}]{\begin{tikzpicture}
    \clip (0,0) rectangle (5.0,5.0);
    \Vertex[x=1.400,y=0.300,size=0.5]{5}
    \Vertex[x=1.400,y=1.400,size=0.5,label=$A$,color=red,opacity=0.7]{6}
    \Vertex[x=1.400,y=2.500,size=0.5,label=$u$]{7}
    \Vertex[x=1.400,y=3.600,size=0.5]{8}
    \Vertex[x=1.400,y=4.700,size=0.5]{9}
    \Vertex[x=2.500,y=0.300,size=0.5]{10}
    \Vertex[x=2.500,y=1.400,size=0.5]{11}
    \Vertex[x=2.500,y=2.500,size=0.5,label=$v$]{12}
    \Vertex[x=2.500,y=3.600,size=0.5]{13}
    \Vertex[x=2.500,y=4.700,size=0.5]{14}
    \Vertex[x=3.600,y=0.300,size=0.5]{15}
    \Vertex[x=3.600,y=1.400,size=0.5]{16}
    \Vertex[x=3.600,y=2.500,size=0.5,label=$w$]{17}
    \Vertex[x=3.600,y=3.600,size=0.5,label=$B$,color=red,opacity=0.7]{18}
    \Vertex[x=3.600,y=4.700,size=0.5]{19}
    \Edge[,color={200,200,200},bend=0,RGB](5)(10)
    \Edge[,color={200,200,200},bend=0,RGB](5)(6)
    \Edge[,color={200,200,200},bend=0,RGB](6)(11)
    \Edge[,color={55,55,55},bend=0,RGB](6)(7)
    \Edge[,color={200,200,200},bend=0,RGB](6)(5)
    \Edge[,color={55,55,55},bend=0,RGB](7)(12)
    \Edge[,color={200,200,200},bend=0,RGB](7)(8)
    \Edge[,color={55,55,55},bend=0,RGB](7)(6)
    \Edge[,color={200,200,200},bend=0,RGB](8)(13)
    \Edge[,color={200,200,200},bend=0,RGB](8)(9)
    \Edge[,color={200,200,200},bend=0,RGB](8)(7)
    \Edge[,color={200,200,200},bend=0,RGB](9)(14)
    \Edge[,color={200,200,200},bend=0,RGB](9)(8)
    \Edge[,color={200,200,200},bend=0,RGB](10)(15)
    \Edge[,color={200,200,200},bend=0,RGB](10)(11)
    \Edge[,color={200,200,200},bend=0,RGB](10)(5)
    \Edge[,color={200,200,200},bend=0,RGB](11)(16)
    \Edge[,color={200,200,200},bend=0,RGB](11)(12)
    \Edge[,color={200,200,200},bend=0,RGB](11)(6)
    \Edge[,color={200,200,200},bend=0,RGB](11)(10)
    \Edge[,color={55,55,55},bend=0,RGB](12)(17)
    \Edge[,color={200,200,200},bend=0,RGB](12)(13)
    \Edge[,color={55,55,55},bend=0,RGB](12)(7)
    \Edge[,color={200,200,200},bend=0,RGB](12)(11)
    \Edge[,color={200,200,200},bend=0,RGB](13)(18)
    \Edge[,color={200,200,200},bend=0,RGB](13)(14)
    \Edge[,color={200,200,200},bend=0,RGB](13)(8)
    \Edge[,color={200,200,200},bend=0,RGB](13)(12)
    \Edge[,color={200,200,200},bend=0,RGB](14)(19)
    \Edge[,color={200,200,200},bend=0,RGB](14)(9)
    \Edge[,color={200,200,200},bend=0,RGB](14)(13)
    \Edge[,color={200,200,200},bend=0,RGB](15)(16)
    \Edge[,color={200,200,200},bend=0,RGB](15)(10)
    \Edge[,color={200,200,200},bend=0,RGB](16)(17)
    \Edge[,color={200,200,200},bend=0,RGB](16)(11)
    \Edge[,color={200,200,200},bend=0,RGB](16)(15)
    \Edge[,color={55,55,55},bend=0,RGB](17)(18)
    \Edge[,color={55,55,55},bend=0,RGB](17)(12)
    \Edge[,color={200,200,200},bend=0,RGB](17)(16)
    \Edge[,color={200,200,200},bend=0,RGB](18)(19)
    \Edge[,color={200,200,200},bend=0,RGB](18)(13)
    \Edge[,color={55,55,55},bend=0,RGB](18)(17)
    \Edge[,color={200,200,200},bend=0,RGB](19)(14)
    \Edge[,color={200,200,200},bend=0,RGB](19)(18)
\end{tikzpicture}}
        \subfloat[Map between edges and vectors. \label{subfig:path_chiralities}]{\begin{tikzpicture}
\clip (2, 0) rectangle (4, 5);
\Vertex[y=0.300,x=2.500,size=0.5,label=$A$,color=red,opacity=0.7]{0}
\Vertex[y=1.400,x=2.500,size=0.5,label=$u$]{1}
\Vertex[y=2.500,x=2.500,size=0.5,label=$v$]{2}
\Vertex[y=3.600,x=2.500,size=0.5,label=$w$]{3}
\Vertex[y=4.700,x=2.500,size=0.5,label=$B$,color=red,opacity=0.7]{4}
\Edge[,lw=0.7, position=right,label=$\ket*{A, c_{A}^{p}}$,bend=-8.531,color={55,55,55},Direct,RGB](0)(1)
\Edge[,lw=0.7,position=left,label=$\ket*{v, c_{uA}}$,bend=-8.531,color={55,55,55},Direct,RGB](1)(0)
\Edge[,lw=0.7,position=right,label=$\ket*{v, c_{uv}}$,bend=-8.531,color={55,55,55},Direct,RGB](1)(2)
\Edge[,lw=0.7,position=left,label=$\ket*{u, c_{vu}}$, color={55,55,55},bend=-8.531,Direct,RGB](2)(1)
\Edge[,lw=0.7,position=right,label=$\ket*{u, c_{vw}}$,bend=-8.531,color={55,55,55},Direct,RGB](2)(3)
\Edge[,lw=0.7,position=left,label=$\ket*{w, c_{wv}}$,bend=-8.531,color={55,55,55},Direct,RGB](3)(2)
\Edge[,lw=0.7,position=right,label=$\ket*{w, c_{wB}}$,bend=-8.531,color={55,55,55},Direct,RGB](3)(4)
\Edge[,lw=0.7,position=left,label=$\ket*{B, c_{B}^{p}}$,bend=-8.531,color={55,55,55},Direct,RGB](4)(3)
\Edge[,loopsize=0.5cm,lw=0.7,position=right,label=$\ket*{A, c_{A}}$,loopposition=330,color={55,55,55},Direct,RGB](0)(0)
\end{tikzpicture}} \hfill
    \end{centering}
    \caption{Notation for edges exemplified in a grid graph. Consider that $A$ and $B$ are two nodes  connected in a 2D grid network. (\subref{subfig:path_selection}) $p$ is a path connecting $A$ and $B$ with hop-distance $4$ traversed by the walker. (\subref{subfig:path_chiralities}) Each edge on the path corresponds to a vector in $\mathcal{H}_W$, which appear in the walker wavefunction throughout movement. The degrees of freedom are defined such that $\ket{x, c_{xy}}$ represents edge $(x, y)$. As an example, the flip-flop operator specified in \eqref{eq:flipflop} maps $\ket{v, c_{vu}} \to \ket{u, c_{uv}}$, while the operator $C_u$ defined in terms of  \eqref{eq:inner_coin} maps $\ket{u, c_{uA}} \to \ket{u, c_{uv}}$.}
    \label{fig:routing_grid}
\end{figure}

We define coin and shift operators that evolve the global state from \eqref{eq:entanglement_operator} to \eqref{eq:cnot_state}. The quantum walk evolution is restricted to neighbor locality, such that, to have the state given in \eqref{eq:cnot_state} at time $t$, all of the wavefunction at time $t - 1$ must exclusively be a superposition of edges incident to $A$, $B$ and its neighbors. There are many ways to define coin and shift operators with this desired behavior and we consider the case where the quantum walk traverses a single path connecting $A$ and $B$. Some auxiliary definitions and assumptions are required to describe the operators in context. Let $p$ be a path of the network connecting $A$ and $B$ with hop distance $\Delta(A, B)$. We assume that every node knows the network topology and that classical information can be transmitted across the nodes. Recall that the edges of $N'$  are mapped to walker states following the relation
\begin{align}
    & (v, u) \to \ket*{v, {c_{vu}}},
\end{align}
and that the self-loop of node $v$ is mapped to the degree of freedom $c_v$, for all $v$. We refer to the edge that connects $A$ to its neighbor in $p$ as $\ket*{A, c_{A}^{p}}$ and the reverse edge that connects $B$ to its precedent node in $p$ as $\ket*{B, c_{B}^{p}}$. This notation is depicted in Figure \ref{fig:routing_grid} for a 2D-grid network. If some quantum operation is performed between $A$ and $B$ at a given time $t$, the propagation process starts at time $t - \Delta(A, B)$.

The extended shift operator used to route information is fixed and given by
\begin{align}
    & S = S_{f} \otimes I_\mathcal{M}, \label{eq:shif}
\end{align}
where $S_{f}$ is defined in \eqref{eq:flipflop}. The coin operator definition is also time-independent, although it depends on the path $p$ chosen. All operators $C_v$ in \eqref{eq:coin_op_assited} have the form
\begin{align}
    & C_v^{p} = \dyad{c_1}{c_2} + \dyad{c_2}{c_1} + \sum_{\substack{c \in \mathcal{C}_v \\ c \neq c_1, c_2}} \dyad{c} \label{eq:inner_coin}
\end{align}
where $c_1$ and $c_2$ refer to the degrees of freedom that represent the edges incident to $v$ in $p$. Thus, we define $C_v^{p}$ specifying $c_1$ and $c_2$ for the nodes of interest. The unitary $C_A$ has $c_1 = \overline{c_{A}}$ and $c_2 = c_A^{p}$. Let $u$ and $w$ be the neighbors of $v \in p \setminus\{A, B\}$ on the path. $C_v$ has $c_1 = c_{vu}$ and $c_2 = c_{vw}$, representing a permutation between the
edges $(v, u)$ and $(v, w)$ in $p$. Finally, the operator $C_B$ has $c_1 = c_B^{p}$ and $c_2 = c_{B}$. It suffices to set $K_v(t) = I_{\mathcal{M}_v}$ in \eqref{eq:unit_expanded} to perform the desired controlled operation between $a$ and $b$, although it is possible to perform operations controlled by $a$ on the qubits in the intermediate nodes as the walker moves by choosing $K_v(t)$ accordingly.

\begin{figure*}
\begin{centering}
\begin{subfigure}[b]{0.17\textwidth}
    \centering
    \begin{tikzpicture}
\clip (1, 0) rectangle (3.5, 5);
\Vertex[y=0.300,x=2.500,size=0.5,label=$A$,color=red,opacity=0.7]{0}
\Vertex[y=1.400,x=2.500,size=0.5,label=$v$]{1}
\Vertex[y=2.500,x=2.500,size=0.5,label=$u$]{2}
\Vertex[y=3.600,x=2.500,size=0.5,label=$w$]{3}
\Vertex[y=4.700,x=2.500,size=0.5,label=$B$,color=red,opacity=0.7]{4}
\Vertex[x=1.400,y=0.300,size=0.47, label=$z$]{5}
\Edge[,lw=0.7, position=right,bend=-8.531,color={200,200,200},Direct,RGB](0)(1)
\Edge[,lw=0.7,position=left,bend=-8.531,color={200,200,200},Direct,RGB](1)(0)
\Edge[,lw=0.7,position=right,bend=-8.531,color={200,200,200},Direct,RGB](1)(2)
\Edge[,lw=0.7,position=left,color={200,200,200},bend=-8.531,Direct,RGB](2)(1)
\Edge[,lw=0.7,position=right,bend=-8.531,color={200,200,200},Direct,RGB](2)(3)
\Edge[,lw=0.7,position=left,bend=-8.531,color={200,200,200},Direct,RGB](3)(2)
\Edge[,lw=0.7,position=right,bend=-8.531,color={200,200,200},Direct,RGB](3)(4)
\Edge[,lw=0.7,position=left,bend=-8.531,color={200,200,200},Direct,RGB](4)(3)
\Edge[,lw=0.7, label=$\ket*{A, \overline{c_{A}}}$, position={above left = 1.8mm},bend=+8.531,color={55,55,55},Direct,RGB](0)(5)
\Edge[,loopsize=0.5cm,lw=0.7,position=right,label=$\ket*{A, c_{A}}$,loopposition=330,color={55,55,55},Direct,RGB](0)(0)
\end{tikzpicture}
    \caption{Initial state.}
    \label{subfig:path0}
\end{subfigure} \hfill
\begin{subfigure}[b]{0.17\textwidth}
    \centering
    \begin{tikzpicture}
\clip (2, 0) rectangle (3, 5);
\Vertex[y=0.300,x=2.500,size=0.5,label=$A$,color=red,opacity=0.7]{0}
\Vertex[y=1.400,x=2.500,size=0.5,label=$v$]{1}
\Vertex[y=2.500,x=2.500,size=0.5,label=$u$]{2}
\Vertex[y=3.600,x=2.500,size=0.5,label=$w$]{3}
\Vertex[y=4.700,x=2.500,size=0.5,label=$B$,color=red,opacity=0.7]{4}
\Edge[,lw=0.7, position=right,label=$\ket*{A, c_p^A}$,bend=-8.531,color={55,55,55},Direct,RGB](0)(1)
\Edge[,lw=0.7,position=left,bend=-8.531,color={200,200,200},Direct,RGB](1)(0)
\Edge[,lw=0.7,position=right,bend=-8.531,color={200,200,200},Direct,RGB](1)(2)
\Edge[,lw=0.7,position=left, color={200,200,200},bend=-8.531,Direct,RGB](2)(1)
\Edge[,lw=0.7,position=right,bend=-8.531,color={200,200,200},Direct,RGB](2)(3)
\Edge[,lw=0.7,position=left,bend=-8.531,color={200,200,200},Direct,RGB](3)(2)
\Edge[,lw=0.7,position=right,bend=-8.531,color={200,200,200},Direct,RGB](3)(4)
\Edge[,lw=0.7,position=left,bend=-8.531,color={200,200,200},Direct,RGB](4)(3)
\Edge[,loopsize=0.5cm,lw=0.7,position=right,label=$\ket*{A, c_{A}}$,loopposition=330,color={55,55,55},Direct,RGB](0)(0)
\end{tikzpicture}
    \caption{Coin in $A$.}
    \label{subfig:path1}
\end{subfigure} \hfill
\begin{subfigure}[b]{0.17\textwidth}
    \begin{tikzpicture}
\clip (1, 0) rectangle (3, 5);
\Vertex[y=0.300,x=2.500,size=0.5,label=$A$,color=red,opacity=0.7]{0}
\Vertex[y=1.400,x=2.500,size=0.5,label=$v$]{1}
\Vertex[y=2.500,x=2.500,size=0.5,label=$u$]{2}
\Vertex[y=3.600,x=2.500,size=0.5,label=$w$]{3}
\Vertex[y=4.700,x=2.500,size=0.5,label=$B$,color=red,opacity=0.7]{4}
\Edge[,lw=0.7, position=right,bend=-8.531,color={200,200,200},Direct,RGB](0)(1)
\Edge[,lw=0.7,position=left,label=$\ket*{v, c_{vA}}$,bend=-8.531,color={55,55,55},Direct,RGB](1)(0)
\Edge[,lw=0.7,position=right,bend=-8.531,color={200,200,200},Direct,RGB](1)(2)
\Edge[,lw=0.7,position=left, color={200,200,200},bend=-8.531,Direct,RGB](2)(1)
\Edge[,lw=0.7,position=right,bend=-8.531,color={200,200,200},Direct,RGB](2)(3)
\Edge[,lw=0.7,position=left,bend=-8.531,color={200,200,200},Direct,RGB](3)(2)
\Edge[,lw=0.7,position=right,bend=-8.531,color={200,200,200},Direct,RGB](3)(4)
\Edge[,lw=0.7,position=left,bend=-8.531,color={200,200,200},Direct,RGB](4)(3)
\Edge[,loopsize=0.5cm,lw=0.7,position=right,label=$\ket*{A, c_{A}}$,loopposition=330,color={55,55,55},Direct,RGB](0)(0)
\end{tikzpicture}
    \caption{Flip-flop shift}
    \label{subfig:path2}
\end{subfigure} \hfill
\begin{subfigure}[b]{0.17\textwidth}
    \begin{tikzpicture}
\clip (1, 0) rectangle (3, 5);
\Vertex[y=0.300,x=2.500,size=0.5,label=$A$,color=red,opacity=0.7]{0}
\Vertex[y=1.400,x=2.500,size=0.5,label=$v$]{1}
\Vertex[y=2.500,x=2.500,size=0.5,label=$u$]{2}
\Vertex[y=3.600,x=2.500,size=0.5,label=$w$]{3}
\Vertex[y=4.700,x=2.500,size=0.5,label=$B$,color=red,opacity=0.7]{4}
\Edge[,lw=0.7, position=right,bend=-8.531,color={200,200,200},Direct,RGB](0)(1)
\Edge[,lw=0.7,position=left,bend=-8.531,color={200,200,200},Direct,RGB](1)(0)
\Edge[,lw=0.7,position=right,label=$\ket*{v, c_{vu}}$,bend=-8.531,color={55,55,55},Direct,RGB](1)(2)
\Edge[,lw=0.7,position=left, color={200,200,200},bend=-8.531,Direct,RGB](2)(1)
\Edge[,lw=0.7,position=right,bend=-8.531,color={200,200,200},Direct,RGB](2)(3)
\Edge[,lw=0.7,position=left,bend=-8.531,color={200,200,200},Direct,RGB](3)(2)
\Edge[,lw=0.7,position=right,bend=-8.531,color={200,200,200},Direct,RGB](3)(4)
\Edge[,lw=0.7,position=left,bend=-8.531,color={200,200,200},Direct,RGB](4)(3)
\Edge[,loopsize=0.5cm,lw=0.7,position=right,label=$\ket*{A, c_{A}}$,loopposition=330,color={55,55,55},Direct,RGB](0)(0)
\end{tikzpicture}
    \caption{Coins in path.}
    \label{subfig:path3}
\end{subfigure}
\begin{subfigure}[b]{0.17\textwidth}
    \begin{tikzpicture}
\clip (1, 0) rectangle (3, 5);
\Vertex[y=0.300,x=2.500,size=0.5,label=$A$,color=red,opacity=0.7]{0}
\Vertex[y=1.400,x=2.500,size=0.5,label=$v$]{1}
\Vertex[y=2.500,x=2.500,size=0.5,label=$u$]{2}
\Vertex[y=3.600,x=2.500,size=0.5,label=$w$]{3}
\Vertex[y=4.700,x=2.500,size=0.5,label=$B$,color=red,opacity=0.7]{4}
\Edge[,lw=0.7, position=right,bend=-8.531,color={200,200,200},Direct,RGB](0)(1)
\Edge[,lw=0.7,position=left,bend=-8.531,color={200,200,200},Direct,RGB](1)(0)
\Edge[,lw=0.7,position=right,bend=-8.531,color={200,200,200},Direct,RGB](1)(2)
\Edge[,lw=0.7,position=left, color={200,200,200},bend=-8.531,Direct,RGB](2)(1)
\Edge[,lw=0.7,position=right,bend=-8.531,color={200,200,200},Direct,RGB](2)(3)
\Edge[,lw=0.7,position=left,bend=-8.531,color={200,200,200},Direct,RGB](3)(2)
\Edge[,lw=0.7,position=right,bend=-8.531,color={200,200,200},Direct,RGB](3)(4)
\Edge[,lw=0.7,position=left,label=$\ket*{B, c_{B}^{p}}$,bend=-8.531,color={55,55,55},Direct,RGB](4)(3)
\Edge[,loopsize=0.5cm,lw=0.7,position=right,label=$\ket*{A, c_{A}}$,loopposition=330,color={55,55,55},Direct,RGB](0)(0)
\end{tikzpicture}
    \caption{Final state.}
    \label{subfig:path4}
\end{subfigure}
\end{centering}
\caption{Protocol execution in a $5$-by-$5$ grid with a quantum walk through a single path. Dark edges depict vectors which have non-zero wavefunction component in a given step. (\subref{subfig:path0}) The initial state of execution is the state generated by the application of the controlled operation demonstrated in \eqref{eq:entanglement_operator}. The node $z$ is shown for illustration purposes and is not for propagation. (\subref{subfig:path1}) The first coin flip permutes the wavefunction into edges $(A,A)$ and $(A,v)$. (\subref{subfig:path2}) The flip-flop shift exchanges edge $(A, u)$ with edge $(u, A)$, moving the walker while mapping the self-loop edge to itself. (\subref{subfig:path3}) After the first coin flip, all subsequent coin operators work as shift operators inside a node, mapping degrees of freedom in order to propel the walker towards $B$. (\subref{subfig:path4}) After $\Delta(A, B)$ steps, the final wavefunction is a uniform superposition between edges $(A, A)$ and $(B, w)$, which can be used to perform an operation controlled by qubit $a$ located in $A$ with target qubit $b$ located in $B$.}
\label{fig:walk_assisted}
\end{figure*}
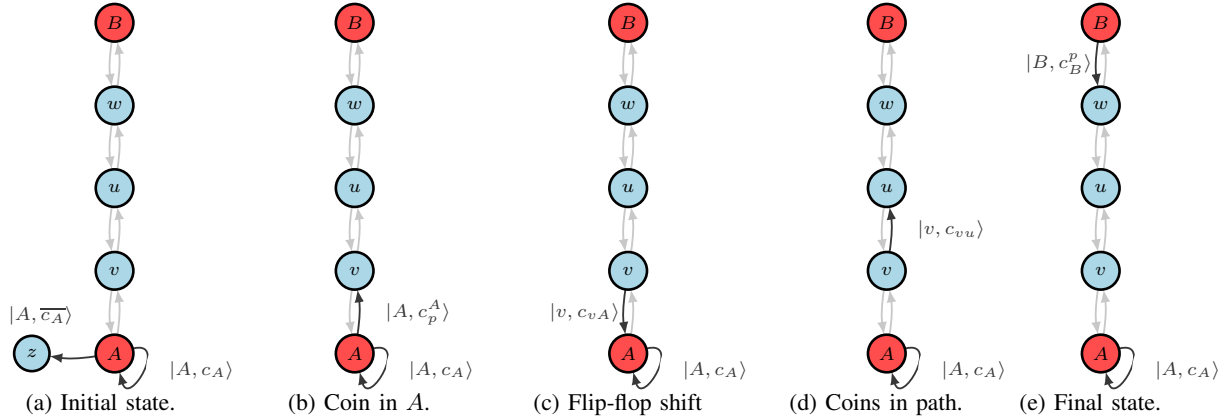

The overall behavior of the walker is straightforward and is illustrated in Figure \ref{fig:walk_assisted}. It is assumed that the operator $O$ defined in \eqref{eq:entanglement_operator} is applied to the system at the beginning of execution.
The first application of the coin operator in $A$ creates a superposition between the states $\ket{A, c_{A}}$ and $\ket{A, c_{A}^{p}}$ that are entangled with $a$. The flip-flop shift allows the wavefunction in $\ket*{A, c_{A}^{p}}$ to propagate along $p$ and ensures that the $\ket*{A, c_{A}}$ component remains in the superposition throughout protocol execution. Each coin flip routes information on nodes internally, ensuring propagation. The net effect of $\Delta(A, B)$ successive applications of $SC$ is the superposition specified in \eqref{eq:cnot_state}.

\subsection{Separating data and control} \label{subsec:separation}

The state in \eqref{eq:cnot_state} is an entangled state between control and data. This implies that a partial trace operation in the walker system does not leave the state of $a$ and $b$ as it should be if the operation was performed without the walker. In order to overcome this problem, the walker evolution is reversed after the controlled operation takes place by applying the inverses of the unitary operators used for propagation. Since the coin and shift operators considered are permutation operators, they are Hermitian unitaries and, thus, are their own inverses. It takes $\Delta(A,B)$ time steps to reverse the walker back to $A$ and to transform the joint state of the system to the form
\begin{align}
    &  \alpha \ket*{A, c_{A}, 0_a, \Psi_b} + \beta \ket*{A, \overline{c_{A}}, 1_a, \overline{\Psi_b}}.
\end{align}
It should be clear that an extended coin operator with interaction of the form given in \eqref{eq:coin_op_assited} and a measurement on the walker produces one of the following separable states
\begin{equation}
    \begin{cases}
        \ket*{A, c_{A}} \otimes (\alpha \ket*{0_a, \Psi_b} + \beta \ket*{1_a, \overline{\Psi_b}}), \\
        \ket*{A, \overline{c_{A}}} \otimes (\alpha \ket*{0_a, \Psi_b} - \beta \ket*{1_a, \overline{\Psi_b}}),
    \end{cases}
\end{equation}
with equal probability. It is straightforward from the separability between the walker system and the qubits $a$ and $b$ that the state after a partial trace operation in the walker system is now equivalent to a CNOT gate, up to a single-qubit $Z$ gate conditioned on the measurement outcome.

Note that moving the walker backwards is only one of several ways to achieve a separable state between the walker and the qubits $a$ and $b$. In a nutshell, any walker dynamics that concentrates the walker's part of the wavefunction into a single network node suffices for this purpose, \textit{e.g.,} propagating the walker wavefunction from both $A$ and $B$ to an intermediate node. Under the assumption of quantum error correction, this backward propagation can occur simultaneously with any further quantum operation that nodes $A$ and $B$ may perform on the qubits and does not impact the latency of the protocol.

\section{Propagation of multiple control signals}

We demonstrated in the previous section how the quantum walk can propagate through a particular path of the network connecting nodes $A$ and $B$. This is easily extended to the traversal of multiple paths as we will show next. In this setting, $k$ walkers can be used to simultaneously perform operations controlled by $a$ with $k$ target qubits $b_j$ located in nodes $B_j$, for $j \in \{0, \ldots, k - 1\}$. This protocol execution translates to the parallel application of $k + 1$ one-qubit gates controlled by $a$ in terms of distributed quantum computation.

The analysis carried for a single walker system extends to this case by considering a set of paths $\mathcal{P}$ such that, for every $j$, the path $p_j \in \mathcal{P}$ connects $A$ and $B_j$. The path $p_j$ defines extended coin operators for the $j$-th walker following \eqref{eq:inner_coin}. By carefully choosing the qubits controlled by each walker, so that each walker interacts with a unique set of qubits, it is possible to route $k = |\mathcal{P}|$ walkers concurrently. This is a trivial extension to the single walker case because the separability of individual coins for each walker yield the $k$-walker coin operator for the set of paths $\mathcal{P}$ to be of the form
\begin{align}
    & C^{\mathcal{P}}(t) = \bigotimes_{j = 0}^{k - 1}C^{p_j}(t). \label{eq:sep_coin}
\end{align}
Since the memories affected by walker $j$ are unique, for every $j \in \{0, \ldots, T - 1\}$, each walker can be seen as a parallel control signal working on distinct qubits. Note that, even though the walker evolution is performed by separable extended coin operators and shifts, the application of the controlled operation $O$ described in \eqref{eq:data_control} before propagation implies that the walkers are maximally entangled with each other and with qubit $a$. Furthermore, the flip-flop shift operator for each walker does not depend on a particular path and the joint operator for all walkers is described by
\begin{align}
    & S^{\mathcal{P}} = \bigotimes_{j = 0}^{k - 1} S, \label{eq:sep_shift}
\end{align}
with $S$ given in \eqref{eq:shif}.

In addition to simultaneous 2-qubit control operations, it is trivial to use multiple walkers departing from $A$ to perform controlled operations in the nodes $B_j$, where the target qubit $b_j$ in $B_j$ is controlled by qubit $a_j$ in $A$. The fundamental difference for this case is that the $O$ operator described in \eqref{eq:entanglement_operator} is controlled by qubit $a_j$ when applied to the $j$-th walker, generating a separable state among the walkers.

\section{Quantum walk propagation and entanglement distribution}

As described previously, the operators defined in \eqref{eq:coin_op_assited} and \eqref{eq:data_control} allow for universal distributed quantum computation among nodes of the network. In this sense, the movement of the walker introduces entanglement across nodes. Given this is the case, it is suitable to consider entanglement distribution protocols in terms of the quantum walk control plane. In this section, we apply the quantum walk control protocol to recover the behavior of two entanglement distribution protocols. In particular, the recovered protocols distribute maximally entangled \ac{GHZ} states across network nodes. We demonstrate, for both protocols, how the entanglement introduced by the walker can be used to produce the same entangled states.

Superposition of the walker wavefunction allows for the creation of a Bell state between $a$ and $b$ as follows. Assume that the joint state of the system is at time $t$ described by
\begin{align}
    & \ket{\Psi(t)} = \frac{1}{\sqrt{2}} (\ket{A, c_A, 0_a, 0_b} + \ket{B, c_B, 0_a, 0_b}),
\end{align}
which is a separable state between the walker and the qubits $a$ and $b$. By setting $U_V = X_v \otimes I$ and $C_V(t) = I$ for $V \in \{A, B\}$ in \eqref{eq:coin_op_assited}, the extended coin operator $C(t)$ is
\begin{align}
    & C(t)\ket{\Psi(t)} = \frac{1}{\sqrt{2}} (\ket{A, c_A, 1_a, 0_b} + \ket{B, c_B, 0_a, 1_b}),
\end{align}
which is an entangled state between $a$ and $b$. The entanglement between $a$, $b$, and the walker system can be removed by transmitting the walker back to $A$ as explained in Section \ref{subsec:separation}. Note that only operators of the form prescribed in \eqref{eq:coin_op_assited} are needed because there is no control dependent on the state of either $a$ or $b$.

A \ac{GHZ} generalizes Bell states to multiple qubits. A $k$-qubit GHZ projection is a von-Neumann measurement operation in the GHZ basis and a Bell state measurement (BSM) is a 2-qubit GHZ projection. 

\subsection{Multi-path entanglement distribution with BSMs}

\begin{figure}
    \centering
        \subfloat[Multi-path propagation.\label{subfig:mp}]{\begin{tikzpicture}
\clip (0,0) rectangle (5.0, 5.0);
\Vertex[x=0.300,y=0.300,size=0.47]{0}
\Vertex[x=0.300,y=1.400,size=0.47]{1}
\Vertex[x=0.300,y=2.500,size=0.47]{2}
\Vertex[x=0.300,y=3.600,size=0.47]{3}
\Vertex[x=0.300,y=4.700,size=0.47]{4}
\Vertex[x=1.400,y=0.300,size=0.47]{5}
\Vertex[x=1.400,y=1.400,size=0.47,label=A,color=red,opacity=0.7]{6}
\Vertex[x=1.400,y=2.500,size=0.47]{7}
\Vertex[x=1.400,y=3.600,size=0.47]{8}
\Vertex[x=1.400,y=4.700,size=0.47]{9}
\Vertex[x=2.500,y=0.300,size=0.47]{10}
\Vertex[x=2.500,y=1.400,size=0.47]{11}
\Vertex[x=2.500,y=2.500,size=0.47]{12}
\Vertex[x=2.500,y=3.600,size=0.47]{13}
\Vertex[x=2.500,y=4.700,size=0.47]{14}
\Vertex[x=3.600,y=0.300,size=0.47]{15}
\Vertex[x=3.600,y=1.400,size=0.47]{16}
\Vertex[x=3.600,y=2.500,size=0.47]{17}
\Vertex[x=3.600,y=3.600,size=0.47,label=B,color=red,opacity=0.7]{18}
\Vertex[x=3.600,y=4.700,size=0.47]{19}
\Vertex[x=4.700,y=0.300,size=0.47]{20}
\Vertex[x=4.700,y=1.400,size=0.47]{21}
\Vertex[x=4.700,y=2.500,size=0.47]{22}
\Vertex[x=4.700,y=3.600,size=0.47]{23}
\Vertex[x=4.700,y=4.700,size=0.47]{24}
\Edge[,color={222,222,222},RGB](0)(1)
\Edge[,color={222,222,222},RGB](0)(5)
\Edge[,color={222,222,222},RGB](1)(2)
\Edge[,color={222,222,222},RGB](1)(6)
\Edge[,color={222,222,222},RGB](2)(3)
\Edge[,color={222,222,222},RGB](2)(7)
\Edge[,color={222,222,222},RGB](3)(4)
\Edge[,color={222,222,222},RGB](3)(8)
\Edge[,color={222,222,222},RGB](4)(9)
\Edge[,color={222,222,222},RGB](5)(6)
\Edge[,color={222,222,222},RGB](5)(10)
\Edge[,color={54,54,54},RGB](6)(7)
\Edge[,color={54,54,54},RGB](6)(11)
\Edge[,color={54,54,54},RGB](7)(8)
\Edge[,color={222,222,222},RGB](7)(12)
\Edge[,color={222,222,222},RGB](8)(9)
\Edge[,color={54,54,54},RGB](8)(13)
\Edge[,color={222,222,222},RGB](9)(14)
\Edge[,color={222,222,222},RGB](10)(11)
\Edge[,color={222,222,222},RGB](10)(15)
\Edge[,color={222,222,222},RGB](11)(12)
\Edge[,color={54,54,54},RGB](11)(16)
\Edge[,color={222,222,222},RGB](12)(13)
\Edge[,color={222,222,222},RGB](12)(17)
\Edge[,color={222,222,222},RGB](13)(14)
\Edge[,color={54,54,54},RGB](13)(18)
\Edge[,color={222,222,222},RGB](14)(19)
\Edge[,color={222,222,222},RGB](15)(16)
\Edge[,color={222,222,222},RGB](15)(20)
\Edge[,color={54,54,54},RGB](16)(17)
\Edge[,color={222,222,222},RGB](16)(21)
\Edge[,color={54,54,54},RGB](17)(18)
\Edge[,color={222,222,222},RGB](17)(22)
\Edge[,color={222,222,222},RGB](18)(19)
\Edge[,color={222,222,222},RGB](18)(23)
\Edge[,color={222,222,222},RGB](19)(24)
\Edge[,color={222,222,222},RGB](20)(21)
\Edge[,color={222,222,222},RGB](21)(22)
\Edge[,color={222,222,222},RGB](22)(23)
\Edge[,color={222,222,222},RGB](23)(24)
\end{tikzpicture}}
    
    \centering
        \subfloat[Spanning tree propagation. \label{subfig:tree}]{\begin{tikzpicture}
\clip (0,0) rectangle (5.0,5.0);
\Vertex[x=0.300,y=0.300,size=0.47,color=green, opacity=0.4]{0}
\Vertex[x=0.300,y=1.400,size=0.47]{1}
\Vertex[x=0.300,y=2.500,size=0.47]{2}
\Vertex[x=0.300,y=3.600,size=0.47]{3}
\Vertex[x=0.300,y=4.700,size=0.47,color=green, opacity=0.4]{4}
\Vertex[x=1.400,y=0.300,size=0.47]{5}
\Vertex[x=1.400,y=1.400,size=0.47,label=A,color=red,opacity=0.7]{6}
\Vertex[x=1.400,y=2.500,size=0.47,label=$u_2$]{7}
\Vertex[x=1.400,y=3.600,size=0.47]{8}
\Vertex[x=1.400,y=4.700,size=0.47]{9}
\Vertex[x=2.500,y=0.300,size=0.47,label=$u_1$]{10}
\Vertex[x=2.500,y=1.400,size=0.47]{11}
\Vertex[x=2.500,y=2.500,size=0.47]{12}
\Vertex[x=2.500,y=3.600,size=0.47]{13}
\Vertex[x=2.500,y=4.700,size=0.47]{14}
\Vertex[x=3.600,y=0.300,size=0.47]{15}
\Vertex[x=3.600,y=1.400,size=0.47]{16}
\Vertex[x=3.600,y=2.500,size=0.47]{17}
\Vertex[x=3.600,y=3.600,size=0.47,label=B,color=red,opacity=0.7]{18}
\Vertex[x=3.600,y=4.700,size=0.47]{19}
\Vertex[x=4.700,y=0.300,size=0.47,color=green, opacity=0.4]{20}
\Vertex[x=4.700,y=1.400,size=0.47,color=green, opacity=0.4]{21}
\Vertex[x=4.700,y=2.500,size=0.47,color=green, opacity=0.4]{22}
\Vertex[x=4.700,y=3.600,size=0.47,color=green, opacity=0.4]{23}
\Vertex[x=4.700,y=4.700,size=0.47,color=green, opacity=0.4]{24}
\Edge[,color={55,55,55},RGB](0)(1)
\Edge[,color={222,222,222},RGB](0)(5)
\Edge[,color={55,55,55},RGB](1)(2)
\Edge[,color={55,55,55},RGB](1)(6)
\Edge[,color={55,55,55},RGB](2)(3)
\Edge[,color={222,222,222},RGB](2)(7)
\Edge[,color={55,55,55},RGB](3)(4)
\Edge[,color={222,222,222},RGB](3)(8)
\Edge[,color={222,222,222},RGB](4)(9)
\Edge[,color={55,55,55},RGB](5)(6)
\Edge[,color={55,55,55},RGB](5)(10)
\Edge[,color={55,55,55},RGB](6)(7)
\Edge[,color={55,55,55},RGB](6)(11)
\Edge[,color={55,55,55},RGB](7)(8)
\Edge[,color={55,55,55},RGB](7)(12)
\Edge[,color={55,55,55},RGB](8)(9)
\Edge[,color={55,55,55},RGB](8)(13)
\Edge[,color={55,55,55},RGB](9)(14)
\Edge[,color={222,222,222},RGB](10)(11)
\Edge[,color={55,55,55},RGB](10)(15)
\Edge[,color={222,222,222},RGB](11)(12)
\Edge[,color={55,55,55},RGB](11)(16)
\Edge[,color={222,222,222},RGB](12)(13)
\Edge[,color={55,55,55},RGB](12)(17)
\Edge[,color={222,222,222},RGB](13)(14)
\Edge[,color={55,55,55},RGB](13)(18)
\Edge[,color={55,55,55},RGB](14)(19)
\Edge[,color={222,222,222},RGB](15)(16)
\Edge[,color={55,55,55},RGB](15)(20)
\Edge[,color={222,222,222},RGB](16)(17)
\Edge[,color={55,55,55},RGB](16)(21)
\Edge[,color={222,222,222},RGB](17)(18)
\Edge[,color={55,55,55},RGB](17)(22)
\Edge[,color={222,222,222},RGB](18)(19)
\Edge[,color={55,55,55},RGB](18)(23)
\Edge[,color={55,55,55},RGB](19)(24)
\Edge[,color={222,222,222},RGB](20)(21)
\Edge[,color={222,222,222},RGB](21)(22)
\Edge[,color={222,222,222},RGB](22)(23)
\Edge[,color={222,222,222},RGB](23)(24)
\end{tikzpicture}}
    
    \caption{Recovering protocol behavior with quantum walks. Protocols that rely on GHZ projections apply entanglement swapping to distribute entanglement between $A$ and $B$. Entanglement swapping is non-unitary, although the overall effect of applying a GHZ projection is a maximally entangled state between qubits. The quantum walk approach recovers this behavior by generating GHZ states unitarily. (\subref{subfig:mp}) A multi-path BSM protocol is recovered by sending separable walkers into each edge-disjoint path used in a round of the protocol, denoted in the figure by shaded edges. Each path forms an independent GHZ state between all qubits in the path. (\subref{subfig:tree}) The GHZ projection protocol is recovered by sending entangled quantum walks to each leaf $l_i$ of a spanning tree of the network rooted in $A$, represented in the figure by green nodes. All qubits in the network are, in this case, in a GHZ state.}
    \label{fig:rec_protocols}
\end{figure}
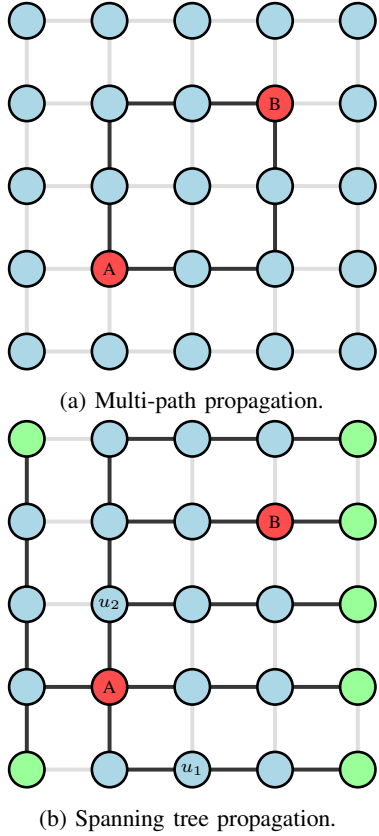

We start with the multi-path protocol defined in \cite{pant2019routing}. Assuming global link state knowledge is available, it suffices to choose the same paths that a complete execution of the multi-path protocol uses to propagate entanglement. Assume that $P = \{p_0, \ldots, p_{k - 1}\}$ is the set of paths used. In this protocol, every path $p_j$ used to distribute entanglement yields a set of $2 (|p_j| - 1)$ qubits in a GHZ state, which comes from the net effect of performing $|p_j| - 2$ BSMs in the intermediate nodes. Moreover, the protocol assumes a network model where every quantum channel in the network is mapped in a one-to-one fashion with pairs of qubits in the nodes incident to that channel. The logical behavior of this protocol is recovered by sending one quantum walker from $A$ through each path $p_j$ such that, once it arrives at $B$, every qubit incident to the quantum channels in the path are in the same GHZ state. Thus, consider that all qubits in the network start in the $\ket{0}$ state. Initialize $k$ walkers in state $\ket{A, c_{A}}$. The operator $C_A^{j}(0)$ for the $j$-th walker can be chosen as any unitary so that
\begin{align}
    & \ket{A, c_{A}} \to \frac{1}{\sqrt{2}} (\ket{A, c_{A}} + \ket{A, c_A^{p_j}}), \label{eq:state_map}
\end{align}
while the interaction operator $K_A^{j}(0)$ is taken to be the identity operator. The state of the system after the application of the extended operator $C(0)$ is
\begin{align}
    & C(0)\ket{\Psi(0)} = \bigotimes_{j} \frac{1}{\sqrt{2}}(\ket*{A, c_{A}} + \ket*{A, c_{A}^{p_j}}).
\end{align}
Protocol execution is performed by the operators defined in \eqref{eq:sep_shift} and \eqref{eq:sep_coin} as specified in the previous section, although the operators $K_v$ are now defined to generate entanglement in the nodes as the walker passes by.
Let $q_{vj0}$ and $q_{vj1}$ denote the pair of qubits inside node $v \in p_j \setminus \{A, B\}$ that need to be entangled together by walker $j$. The operator $U_v^{j}$ in equation \eqref{eq:unit_expanded} is defined for walker $j$ by specifying
\begin{align}
    & K_v(t) = X_{qj0} \otimes X_{qj1} \bigotimes_{q \neq q_{vj0}, q_{vj1}} I_q
\end{align}
for all $t$, performing an $X$ gate on qubits $q_{v0}^{j}$ and $q_{v1}^{j}$ controlled by the position of the $j$-th walker. Since paths are all edge-disjoint, every qubit in the graph spanned by $\mathcal{P}$ interacts with exactly one walker under the application of the generalized coin operator. For walker $j$, the interaction operator with qubit $a_j$ is given by
\begin{equation}
    K_A(t) =
        \begin{cases}
            X_{a_{j}} \otimes I, \text{ when } t = 1,\\
            I_{a_{j}} \otimes I, \text{ when } t \neq 0,
        \end{cases}
\end{equation}
where the identity operator after the tensor product symbol acts on all qubits in $A$ except $a_j$. Considering that $u_j$ is the first neighbor of $A$ in the path $p_j$, the application of the evolution operator $S(0)C(0)$ gives
\begin{align}
    & \ket{\Psi(1)} = \bigotimes_{j} \frac{1}{\sqrt{2}} (\ket*{A, c_{A}} + \ket*{u_j, c_{u_{j}A}}),
\end{align}
where the data qubits are omitted for simplicity. $S(1)C(1)$ generates an entangled state between $a_j$ and the qubits $q_{uj0}$ and $q_{uj1}$ and moves the walker to the neighbors that are $2$ hops away from $A$. Every application of $S(t)C(t)$ for $t > 1$ increases by two the number of qubits entangled with walker $j$ such that, when $t = \Delta(A, B)$, the walker $j$ is in an entangled state with the qubits in $p_j$, $a_j$ and $b_j$. Figure \ref{fig:rec_protocols}(\subref{subfig:mp}) depicts protocol execution for a $5$-by-$5$ grid network. In this case, $P$ is formed by two shortest-paths in the grid, i.e $K = 2$, and the entire evolution of the walkers takes four steps.

\subsection{Distributing GHZ states across network nodes}

GHZ projectors are used in \cite{patil2020entanglement} to improve the multi-path BSM protocol described in \cite{pant2019routing}. The logical behavior of the GHZ distribution protocol can be recovered in the quantum walk protocol by slightly modifying the description presented for the multi-path BSM protocol. In this protocol, every qubit in a node used to distribute entanglement is part of the final GHZ state achieved. This is in clear contrast with the BSM multi-path protocol, where each path defines an independent GHZ state. The first modification is that, instead of preparing separable walkers with coins that maps states following the form given in \eqref{eq:state_map}, the $k$ walkers need to be entangled in the state
\begin{align}
    & \frac{1}{\sqrt{2}} (\ket{A, c_{A}, \ldots, A, c_{A}} + \ket{A, c_A^{p_0}, \ldots, A, c_A^{p_j}}).
\end{align}
Note that this behavior can be obtained with an interaction operator of the form \eqref{eq:data_control} and a coin operator in the form specified by \eqref{eq:inner_coin}. Secondly, instead of routing walkers through edge-disjoint paths, the walkers traverse a tree of the network. More concretely, let $\mathcal{T}$ be a spanning tree of the lattice, rooted at $A$. $k$ is now the number of leaves in $\mathcal{T}$ and $\mathcal{P}$ is the set of paths in the tree connecting $A$ with leaves. For any node $v$, the interaction operator between the walker and the data qubits is
\begin{align}
    & K_v = \bigotimes_{q_v} X_{qv}^{\frac{1}{W_v}},
\end{align}
where $W_v$ is the number of walkers that pass through $v$, i.e the number of leaves connected to $A$ by $v$ in the tree. Note that $K_v$ is the same for every walker. The exponent $\frac{1}{W_v}$ is required because more than one walker may pass through the same vertex, and operate on the same set of qubits. Since paths are taken from a spanning tree, each walker can take a particular time interval to traverse its path. Thus, the time  necessary to entangle all of the necessary qubit in the network and remove the entanglement between the walkers and the data qubits is on the order of $\mathcal{O}(\max_j |p_j|)$, where $|p_j|$ denotes the hop distance of path $p_j$. Again, protocol execution for the generic GHZ projector case is exemplified for a $5$-by-$5$ grid in Figure \ref{fig:rec_protocols}(\subref{subfig:tree}). In this case, $k = 7$ and the whole process takes seven time steps to complete, since the size of the largest path in the tree is seven. For nodes $u_1$ and $u_2$, the value of $W$ is one and three respectively.
\section{Conclusion}

The quantum walk protocol proposed in this article provides a logical description for a network control plane capable of performing universal distributed quantum computing. The description abstracts the implementation of the quantum walker system, as well as the implementation of quantum operations in the network nodes. We consider that quantum error correction yields the application of perfect unitary operators. The key idea that the protocol builds upon is the use of a quantum walker system as a quantum control signal that propagates through the network one hop a time. In spite of abstracting physical implementations, the propagation of the walker stipulates latency constraints for the protocol. A generic controlled operation between a qubit in node $A$ with a qubit in node $B$ demands $\mathcal{O}(\Delta(A, B))$ steps of walker evolution. In the context of a possible physical realization of such control system, this latency constraints translates directly to the physical distance between nodes $A$ and $B$. The description of the protocol in the logical setting also masks the effects of walker propagation in the fidelity of distributed operations. When considering imperfect operators, the fidelity of the final outcome is bounded by the fidelity of the coin and shift operators in the quantum walk system. Throughout this manuscript, the propagation of a quantum walk across the network depended on a particular path of the network to be traversed. This has the clear requirement that every node in the network knows the network topology and that the operators used can be defined by the exchange of classical messages between the nodes. The protocol described was used to represent entanglement distribution protocols defined in the literature in terms of quantum control information exchanged between the nodes in the network. This result highlights connections between the proposed protocol and entanglement distribution protocols.

There are two clear directions for future work considering our results. The first relates to the investigation of a network implementation for a quantum walk system. Such implementation would allow for a realistic characterization of quantities like fidelity and latency. The second point is the description of control exclusively with quantum information. In this setting, nodes would transmit a quantum state containing both the state of the walker and the target node to which control information must be transmitted to.


{\em Acknowledgments}---This research was supported in part by the NSF grant CNS-1955834, NSF-ERC Center for Quantum Networks grant EEC-1941583 and by the National Science Foundation to the Computing Research Association for the CIFellows 2020 Program.
    
\bibliography{references}
\bibliographystyle{unsrt}

\end{document}